\documentclass[12pt]{article}
\begin{document}
{\LARGE
Comment on ``Noncommutativity as a Possible 
Origin of the Ultrahigh-energy Cosmic Ray and the TeV Photon Paradoxes" }
\phantom{aaa} 

\centerline{\Large Cosmas Zachos }  
High Energy Physics Division,
Argonne National Laboratory, Argonne, IL 60439-4815, USA \\
\phantom{a}\qquad  \qquad \qquad  {\sl zachos@anl.gov}

\begin{abstract} 
A Lorentz-noninvariant modification of the kinematic dispersion 
law was proposed in Ref \cite{commentee}, claimed to be derivable 
from from $q$-deformed noncommutative theory, and argued to evade
ultrahigh energy threshold anomalies (trans-GKZ-cutoff  cosmic rays and 
TeV-photons) by raising the respective thresholds. 
It is pointed out that such dispersion laws 
do not follow from deformed oscillator systems, and the proposed
dispersion law is invalidated by tachyonic propagation, as well as  
photon instability, in addition to the process considered. 
\end{abstract}
\bigskip

\hrule

\bigskip

Reported anomalous cosmic threshold events serve to constrain 
hypothesized violations of 
Lorentz invariance \cite{coleman,amelino,proth1,kifune}.
E.g., observations of roughly isotropic cosmic rays
with $p\geq 4\times 10^{19}$eV,  which are above the
Greisen-Zatsepin-Kuz'min (GZK) threshold, or else, detection of up to 
20 TeV photons from the active galaxy (blazar) Markarian 501.
The GZK threshold for $ p+ \gamma \rightarrow p+ \pi$,  ensures absorption,
within about 50 Mpc, of cosmic rays such as protons  with energies above 
$p_{GKZ}=((m_p+m_\pi)^2-m_p^2)/(4k)$, where $k$ is an active energy 
of available cosmic microwave background photons,  $\sim 10^{-3}-10^{-4} eV$.
Similarly, reported 20 TeV photons from Markarian 501 
(at a distance of 157 Mpc, much longer than the mean free path for photons 
cut off by collisions with infrared photons to pair-produce, 
$\gamma + \gamma \rightarrow e^{+} + e^{-}$) should have been absorbed: the 
cutoff can effectively set in at about $p^{\gamma}=m_e^2/k$, where $k$ is 
an active energy of universal IR background photons, $\sim 10^{-1} eV$,
thought to result out of star formation in the early universe \cite{proth1}.
These thresholds might be evaded by small deformations of Lorentz 
kinematics, for instance by raising them to higher 
energies which allow for the reportedly observed 
events \cite{coleman,amelino,proth1,kifune}. 

One such kinematic energy-momentum dispersion law was proposed 
in ref \cite{commentee}, namely,
\begin{equation}
\label{urfunkydispersion}
E=\frac{\omega}{2}\left(\left[\frac{\sqrt{m^2+p^2}}{\omega}
-\frac{1}{2}\right]_q+\left[\frac{\sqrt{m^2+p^2}}{\omega}
+\frac{1}{2}\right]_q\right)=  
\frac{\omega}{2} \left[\frac{\sqrt{m^2+p^2}}{\omega/2}\right]_{\sqrt{q}},
\end{equation}
where 
\begin{equation} 
[x]_q\equiv\frac{q^x-q^{-x}}{q-q^{-1}}~~.
\end{equation}
The deformation parameter $q$ is taken to be close to unity,
so that, for $\kappa\equiv \ln q$, the standard Lorentz kinematic 
dispersion is modified,
\begin{equation}
\label{funkydispersion}
E=\frac{\omega}{2} \frac{\sinh (\kappa\sqrt{m^2+p^2}/\omega )}
{\sinh (\kappa /2)  }   =\sqrt{m^2+p^2} \left ( 1 + 
\frac{ \kappa^2}{24} \left (\frac{ 4(m^2+p^2)}{\omega^2}-1 \right )\right ) 
+ O(\kappa^4).
\end{equation}
The other parameter, $\omega/2$, is the universal energy scale needed for 
a nonlinear law, suggested in ref \cite{commentee} to be a fraction of an MeV.
(NB. \ The actual rest mass of the particle is finitely ``renormalized" to 
$E(p=0)$, i.e., $M=\omega ~{\sinh (\kappa m /\omega )} 
/(2~ {\sinh (\kappa /2)  }) $.)

This law, for both fermions and bosons, was argued in ref \cite{commentee} 
to follow from standard deformed oscillator systems \cite{iwata}.
It is pointed out that this law cannot follow from deformed oscillator systems, 
and that it does not address the threshold anomaly paradoxes satisfactorily.

The starting point of ref \cite{commentee} 
is the standard q-fermion oscillator hamiltonian \cite{ng}, 
\begin{equation}
\label{fermi-H-n}
\hat{H}_q=\frac{\omega}{2}\left([\hat{N}]_q-[\widehat{1-N}]_q\right),
\end{equation}
where the number operator $\hat{N}$ counts creation and annihilation 
operators for self-excuding fermions (or deformed fermions, 
likewise self-excluding \cite{ng}), and has eigenvalues 1 or 0 only. 
It is thus idempotent, $\hat{N}^2= \hat{N}$, and, as a consequence, 
$q^{\hat{N}}=1+\hat{N} (q-1)$. Consequently, it has been broadly appreciated, 
for a while \cite{jing}, that the q-hamiltonian for fermions
(\ref{fermi-H-n}) is {\em no different} than the free fermion 
oscillator hamiltonian
limit of it ($q=1$), viz.  $\hat{H}={\omega}(\hat{N} -1/2)$, despite 
appearances to the contrary.

Thus, no nontrivial inference may be drawn from the q-fermion 
hamiltonian (\ref{fermi-H-n}), let alone the proposed violation of 
Lorentz invariance. 

Beyond fermions, ref \cite{commentee} also addresses q-bosons,
through the popular Biedenharn-McFarlane hamiltonian,
for bosonic q-oscillators commuting with each other, and integer-valued
$\hat{N}$, 
\begin{equation}
\label{qH}
\hat{H}_q=\frac{\omega}{2}\left ([\hat{N}]_q+[\hat{N}+1
]_q\right)=
\frac{\omega}{2}\left(\left [{\hat{h}\over\omega} -{1\over 2} \right ]_q
+\left [{\hat{h}\over\omega} +{1\over 2}\right ]_q\right ),
\end{equation}
where $\hat{h}$ is the undeformed free hamiltonian. 
This is interacting \cite{VokosZachos}, not free, as its
spectrum is not linearly spaced. Conceivably, it specifies some nonlinear 
medium of sorts, and leads to a modification of the 
Planck distribution \cite{martin} (which  has not been fully 
evaluated analytically; nevertheless, its critical properties are
different than those of a boson gas, \cite{monteiro,salerno}). 
Such distributions might allow constraining 
the parameters $\kappa$ and $\omega$ through comparison with cosmic microwave 
background spectral data, but would evidently provide poor constraints.
(They might also affect the average impacted photon energies assumed in GKZ 
threshold analyses, but not drastically.) 

In second-quantizing noncommutative scalar fields \cite{carmona}, one 
normally assembles an infinity of oscillators of different momenta $p$, each 
of which, in fact, could be related to the above, through suitable deformations 
\cite{poly}. 
One then works out the dependence of the momentum-dependent coefficient of
the plain oscillators in the free hamiltonian to obtain a dispersion 
law $E(p)$ in such media, and to then investigate its Lorentz-violating 
implications \cite{carmona}. By contrast, ref \cite{commentee} proposes 
to merely substitute the eigenvalue 
$E=\sqrt{m^2+p^2}$ for the free hamiltonian $\hat{h}$ 
inside the {\em single-mode} (first quantized) interacting 
(\ref{qH}) to produce the dispersion law (\ref{urfunkydispersion}),
claimed to hold for this medium, i.e., to have the nonlinear interaction 
provide a deformation of the kinematics for {\em all} momenta! 
This is plainly unwarranted. 

In conclusion, for both fermions and bosons, deformation hamiltonians do not 
lead to the modified kinematic dispersion 
(\ref{urfunkydispersion},\ref{funkydispersion}).

As an aside, rather than stretch for supporting arguments and motivation, 
one might alternatively consider this dispersion as a mere ad hoc heuristic: 
i.e., simply postulate the dispersion 
(\ref{urfunkydispersion},\ref{funkydispersion}), as a modifier of the 
index of refraction $p/E$ in a hypothetical medium, and 
ask if it raises the cosmic thresholds to address the more severe 
multi-TeV-photon potential paradox. It is argued that it does not succeed.

The dispersion (\ref{funkydispersion}) is plagued by a tachyonic 
region, since the group velocity for $m=0$ is
\begin{equation}
\frac{dE}{dp}= \frac{ \kappa \cosh (\kappa   p/\omega)}{2 \sinh (\kappa /2) }
=1+ \frac{ \kappa^2}{2} ~\left (\frac{p^2}{\omega^2}  -{1\over 12}\right )
+ O(\kappa^4), 
\end{equation}
tachyonic for, e.g., small $\kappa$ and for sufficiently energetic photons,
$p\geq \omega/\sqrt{12}$, well within the parameter ranges utilized in
ref \cite{commentee}. The pitfalls of such superluminal propagation 
are unforgiving to causality \cite{lehnert}. 

The index of refraction is 
\begin{equation}
\frac{p}{E}= \frac{ 2p ~ \sinh (\kappa /2)}{\omega ~ \sinh (\kappa p/\omega)}~,
\end{equation}
smaller than the conventional undeformed value ($\kappa=0$), since sinh (and cosh) 
are positive definite and convex for positive arguments.

Furthermore, the photon in this picture is severely kinematically 
unstable to decay, $\gamma \rightarrow e^{+}+ e^{-}$ \cite{coleman}.
At threshold for decay to an electron-positron pair, the leptons have 
half the momentum $p$ of the photon, so from energy conservation and 
(\ref{funkydispersion}), photons of momentum $p$ decay if 
\begin{equation}
\sinh \left ({ p \kappa \over \omega}\right ) \geq 2\sinh \left (
{\kappa\over \omega}\sqrt{m_e^2 + \frac{p^2}{4}} \right ).
\end{equation}
The corresponding threshold condition for 
$\gamma + \gamma \rightarrow e^{+} + e^{-}$ contemplated in ref 
\cite{commentee}, in the frame where the microwave and IR background is 
essentialy isotropic, should amount to 
\begin{equation}
\sinh \left ({ p \kappa \over \omega}\right )+ 
\sinh \left ({ k \kappa \over \omega}\right ) 
\geq 2\sinh \left ({\kappa \over \omega} \sqrt{m_e^2 
+ \frac{(p-k)^2}{4}} \right  ).
\end{equation}
The undeformed limit ($\kappa=0$) threshold for absorption is  
$p^{\gamma} =  m_e^2/k$, where $k\sim 10^{-1} eV$.
For such $p^{\gamma}\gg m_e \gg k$, both processes above converge to 
each other and both absorb multi-TeV photons by the convexity of the 
hyperbolic sine functions for positive arguments, $\sinh (2x) > 2 
\sinh x$, so such photons do not survive. ~~(Ref \cite{commentee} employs a 
$\kappa /\omega$ ranging from $ 10^{-9} TeV^{-1}$, 
effectively linearizing the formulas, to large values, for which  
the leading exponentials in both of the above formulas dominate the 
hyperbolic functions, and hence 
photons of momentum larger than $2 {\omega \over \kappa} ~\ln 2 \sim 1.39 
{\omega \over \kappa}$ are absorbed.)

\bigskip 

{\it Note Added in Proof:}~~~ F Stecker kindly points out that the 
extragalactic 20 TeV photon ``anomaly" of ref \cite{proth1} is itself 
unwarranted, and may be explained away through proper consideration of 
intergalactic absorption \cite{stecker},
thereby providing accurate confirmation of Lorentz invariance. 
The purported observational absence or displacement of the GKZ cutoff has 
also been seriously controverted \cite{bahcall}.
\bigskip

\noindent{\it Acknowledgement} ~~~
This work was supported by the US Department of Energy, 
Division of High Energy Physics, Contract W-31-109-ENG-38.

\end{document}